\documentclass[aps,prd,showpacs,superscriptaddress,nofootinbib,preprintnumbers,notitlepage]{revtex4-1} 
\usepackage{graphicx}
\usepackage{epsf}
\usepackage{bm}
\usepackage{amsmath}
\usepackage{amsfonts}
\usepackage{comment}
\usepackage{amssymb}
\usepackage{epstopdf}
\usepackage{natbib}
\usepackage{hyperref}
\usepackage{color}
\usepackage{verbatim}
\usepackage{multirow}
\usepackage{bm}
\usepackage{xcolor}
\usepackage{hyperref}
\definecolor{darkblue}{rgb}{0.0, 0.0, 0.55}
\definecolor{darkred}{rgb}{0.55, 0.0, 0.0}
\usepackage{hyperref}
\hypersetup{
    colorlinks=true,
    linkcolor=darkblue,
    citecolor=darkblue,
    urlcolor=darkblue}
    
\makeatletter\let\expandableinput\@@input\makeatother

\begin{document}

\title{One-loop power spectrum corrections in interacting dark energy cosmologies}

\author{Emanuelly Silva}
\email{emanuelly.santos@ufrgs.br}
\affiliation{Instituto de F\'{i}sica, Universidade Federal do Rio Grande do Sul, 91501-970 Porto Alegre RS, Brazil}

\author{Gabriel Hartmann}
\email{gabriel.hartmann@ufrgs.br}
\affiliation{Instituto de F\'{i}sica, Universidade Federal do Rio Grande do Sul, 91501-970 Porto Alegre RS, Brazil}

\author{Rafael C. Nunes}
\email{rafadcnunes@gmail.com}
\affiliation{Instituto de F\'{i}sica, Universidade Federal do Rio Grande do Sul, 91501-970 Porto Alegre RS, Brazil}
\affiliation{Divisão de Astrofísica, Instituto Nacional de Pesquisas Espaciais, Avenida dos Astronautas 1758, São José dos Campos, 12227-010, São Paulo, Brazil}

\begin{abstract}
Interacting dark energy (IDE) models offer a promising avenue to explore possible exchanges of energy and momentum between dark matter and dark energy, providing a dynamical extension of the standard $\Lambda$ cold dark matter ($\Lambda$CDM) paradigm. Such interactions modify the growth of cosmic structures, imprinting distinctive signatures on the matter power spectrum that can be tested through large-scale structure (LSS) observations. In this work, we compute the one-loop corrections to the matter power spectrum in IDE models. We then reinterpret these results within the standard framework of the effective field theory of large-Scale structure (EFTofLSS), which provides a consistent description of mildly nonlinear scales and allows for reliable comparisons with observational data. We investigate two commonly studied forms of the coupling function, $Q$, namely $Q = \xi \mathcal{H} \rho_{\rm m}$ and $Q = \xi \mathcal{H} \rho_{\rm DE}$, and introduce a novel interaction term, $Q = \Gamma \, \rho_{\rm m} \, \rho_{\rm DE} \, \theta_{\rm m}$, characterized by the nonlinear coupling constant $\Gamma$, which links the interaction strength to the velocity divergence of dark matter. This coupling function is proposed to isolate the effects solely of the IDE model on mildly nonlinear scales. Using full-shape (FS) measurements of the galaxy power spectrum from BOSS DR12, we constrain the interaction rate $\Gamma$, the cosmological parameters, and the bias parameters. We find $\Gamma = 0.0039 \pm 0.0082$, which is highly consistent with the $\Lambda$CDM model. This work opens the possibility of testing IDE models at mildly nonlinear scales, potentially providing new insights for this class of models beyond the standard $\Lambda$CDM framework.
\end{abstract}

\maketitle

\section{Introduction}
Cosmological experiments have achieved remarkable gains in both precision and performance over the past decades, driving current data to exhibit exceptionally small statistical uncertainties. When observational error bars were substantially larger, results from different surveys could readily overlap within their uncertainties, naturally supporting concordance among independent measurements. Today, however, the dramatic reduction in these uncertainties has revealed a series of cosmological tensions that challenge the robustness of the standard cosmological model, $\Lambda$CDM. Chief among these is the well-known Hubble tension, a discrepancy exceeding $7\sigma$ between the value of the Hubble constant, $H_0$, inferred from cosmic microwave background (CMB) measurements within the $\Lambda$CDM framework and the higher value obtained from local distance ladder determinations by the SH0ES team~\cite{H0DN:2025lyy} (see \cite{CosmoVerseNetwork:2025alb} for a recent review). Another significant challenge concerns the parameter $S_8 \equiv \sigma_8 \sqrt{\Omega_m/0.3}$, which encodes the amplitude of matter clustering ($\sigma_8$) and the matter density parameter $\Omega_m$. Early weak-lensing surveys reported a tension at the $3\sigma$ level or higher~\cite{KiDS:2020suj,DES:2017qwj}, although more recent KiDS analyses find consistency with $\Lambda$CDM predictions~\cite{Wright:2025xka}. Measurements of $S_8$ from ACT-CMB data likewise align with Planck results~\cite{ACT:2025fju}, and several independent surveys report similarly concordant constraints~\cite{Sailer:2024coh,Chen:2024vvk,Anbajagane:2025hlf,Garcia-Garcia:2024gzy,Sugiyama:2023fzm}. At the same time, full-shape (FS) galaxy clustering analyses with BOSS (Baryon Oscillation Spectroscopic Survey) data continue to report a significant tension, reaching up to $4.5\sigma$ relative to early universe expectations~\cite{Ivanov:2024xgb,Chen:2024vuf}, while additional hints of discrepancy in $S_8$ arise from redshift-space distortion (RSD) measurements~\cite{Nunes_Vagnozzi_2021,Kazantzidis_2018} and other late-time probes~\cite{Karim:2024luk,Dalal:2023olq}.

Addressing these tensions will require not only careful analyses and reanalyses of current and upcoming data within the $\Lambda$CDM framework, but also systematic tests of alternative cosmological models capable of mitigating or resolving such discrepancies. In this context, surveys that probe the large-scale structure (LSS) of the Universe play a particularly crucial role, as they provide complementary and independent information on the distribution and growth of cosmic structures. Recent FS measurements from the first year of Dark Energy Spectroscopic Instrument (DESI) have already achieved unprecedented precision in clustering observables, enabling more rigorous tests of the dark energy sector, providing strong constraints on the sum of neutrino masses, and offering new opportunities to test modified gravity scenarios \cite{DESI:2024hhd, DESI:2025hao, DESI:2024jxi, Ishak:2024jhs, NovellMasot:2025fju, Chudaykin:2025lww}. These results further point toward compelling evidence for a dynamical dark energy equation of state, suggesting possible departures from the cosmological constant paradigm. Such findings underscore the power of LSS observables in probing the dynamics of the Universe and motivate continued efforts from past surveys such as BOSS, ongoing projects like DESI~\cite{DESI:2025kuo,DESI:2025ejh,DESI:2025zpo} and Euclid~\cite{Euclid:2025diy,Euclid:2024yrr}, and forthcoming missions including LSST (Legacy Survey of Space and Time)~\cite{LSST:2008ijt} and the Nancy Grace Roman Space Telescope~\cite{Kessler:2025eib}.

Despite the remarkable progress enabled by current surveys, fully exploiting the potential of present and forthcoming data requires robust theoretical and observational frameworks capable of extracting the maximal information content from each cosmological model, a task that remains highly challenging. In this regard, a broad range of studies has been devoted to this effort, both within the successful $\Lambda$CDM paradigm \cite{Chudaykin:2025aux, Sabogal:2025qhz, Paliathanasis:2025kmg, Elbers:2024dad} and in the context of several alternative scenarios \cite{Sabogal:2025mkp, Zheng:2025owb, Silva:2025twg, Chudaykin:2025gdn, Navone:2025gxr, Li:2025vqt, Smith:2025icl, Akarsu:2025nns, Luciano:2025dhb, Reboucas:2025hue}. Among these, interacting dark sector (IDE) models—postulating nongravitational interactions between dark energy (DE) and dark matter (DM)—continue to be systematically investigated from both theoretical and observational perspectives~\cite{vanderWesthuizen:2025mnw, vanderWesthuizen:2025rip, Zhang:2025dwu, Li:2025muv, Wang:2025znm, Yashiki:2025loj, Cruickshank:2025iig, Pan:2025qwy, Feng:2025mlo, Zhai:2025hfi, Liu:2025vda, Li:2025vuh, Chakraborty:2025syu, Califano:2024xzt, Halder:2024aan, Li:2025owk, Lyu:2025nsd, Bansal:2024bbb, Johnson:2021wou}. These models are particularly appealing because they preserve the empirical successes of $\Lambda$CDM while offering potential pathways to alleviate current cosmological tensions~\cite{DiValentino:2017iww, Sabogal:2024yha, Liu:2023kce, Zhang:2025dwu, Silva:2025hxw}. Previous analyses of IDE effects on LSS have focused primarily on the linear regime, corresponding to very large scales where $\delta \ll 1$ (see, e.g.,~\cite{Li:2013bya, Duniya:2015nva, Valiviita:2008iv, Lucca:2021eqy, Sinha:2022dze, Sharma:2021ivo, Asghari:2020ffe, Liu:2025pxy, Pooya:2024wsq, Sun:2013pda, Duniya:2015nva, Eingorn:2015rma, Yang:2014okp, Goh:2024exx, Archidiacono:2022iuu}). However, it is now well established that the mildly nonlinear regime contains substantial additional information and, crucially, enables the breaking of degeneracies among cosmological parameters. Recent efforts such as those presented in~\cite{Nunes:2022bhn, Silva:2024ift} constitute the first systematic attempts to probe nonlinear scales in interacting dark-sector scenarios within a unified theoretical-observational framework. Specific classes of IDE models have also been investigated at nonlinear scales using $N$-body simulations~\cite{Hashim:2018dek, Maccio:2003yk, Baldi:2008ay}, yielding interesting results into the small-scale phenomenology associated with dark-sector interactions. Several classes of models beyond the standard $\Lambda$CDM cosmology are expected to exhibit significant effects at nonlinear scales. Recently, the calculation of perturbative theoretical predictions beyond $\Lambda$CDM has attracted considerable interest, particularly in the context of coupled DE~\cite{Tudes:2024jpg} and modified gravity models~\cite{Lewandowski:2019txi, Sugiyama:2023tes, Hirano:2022yss, Lewandowski:2016yce}.

As has been well established in the recent literature, IDE scenarios lead to modifications in the continuity and Euler equations, which are key components of the standard perturbation theory (SPT) framework \cite{Bernardeau:2001qr, Scoccimarro:1995if, Scoccimarro:1997st}. Given the growing interest in this class of models within the cosmological community, in this work we compute the one-loop corrections to the matter power spectrum for some of the most widely studied IDE parametrizations available in the literature. Since SPT organizes the evolution of density and velocity fields through a systematic expansion around the linear regime, it provides a natural starting point to assess how IDE interactions alter mode coupling and the growth of structure.

We then interpret these corrections within the framework of the effective field theory of large-Scale structure (EFTofLSS) (for a few reference \cite{Baumann:2010tm, Senatore:2014vja, Carrasco:2013mua, Blas:2015qsi, DAmico:2022ukl, Anastasiou:2025jsy}), which provides a systematic and self-consistent description of structure formation beyond the linear regime. This approach accounts for the influence of small-scale nonlinearities on large-scale observables through a controlled perturbative expansion. A variety of theoretical models have recently been confronted with observations within this formalism \cite{Chudaykin:2025lww, Silva:2025twg, Tsedrik:2025cwc, Zhang:2024thl, Chen:2024vuf, Noriega:2024eyu, Tsedrik:2022cri, Carrilho:2022mon, Simon:2022lde, Simon:2022csv, Shim:2024pce, DESI:2025wzd, Reeves:2025bxc, Toomey:2025yuy, Lu:2025sjg, Spaar:2023his, Lu:2025gki, Bottaro:2023wkd}. Our analysis demonstrates that, at nonlinear scales, IDE-type models require careful theoretical modeling and improved understanding in order to achieve accurate predictions and meaningful comparisons with forthcoming high-precision data.

The structure of this \textit{paper} is as follows. In Sec.~\ref{sec:model}, we present a brief overview of the IDE framework at the background level and discuss its implications for linear perturbation theory. Section~\ref{sec:nl_IDE} examines the nonlinear perturbative regime, focusing on the impact of IDE on the evolution of perturbations, the modification of mode-coupling kernels, and the behavior of interactions that become relevant on mildly nonlinear scales. In Sec.~\ref{sec:1L_IDE}, we analyze the resulting changes in the matter power spectrum, including the one-loop corrections induced by IDE. Section~\ref{sec:constraints_IDE} outlines our analysis pipeline, detailing the observational datasets used to test the proposed framework and the statistical methods employed, followed by a comprehensive discussion of the resulting parameter constraints. Finally, Sec.~\ref{sec:end_IDE} summarizes our main findings and highlights promising directions for future work.

\section{IDE review}
\label{sec:model}
In this section, we review the theoretical framework of IDE models, focusing on their formulation at both the background and perturbation levels.  We present the general form of the modified continuity and Euler equations, emphasizing how the coupling term affects the evolution of DM and DE densities as well as the growth of structure in the linear regime. The discussion follows the standard treatment in the literature and serves as the foundation for the nonlinear analysis developed in the subsequent sections.

In the IDE framework for describing interactions within the dark sector, a parametrization is introduced into the conservation laws, ensuring that while the total energy-momentum tensor remains conserved, the individual contributions of DE and DM do not satisfy independent conservation equations~\cite{Wang:2024vmw}. This condition is expressed as

\begin{equation}
    \sum_{j} \nabla_\mu T_{j}^{\mu\nu} = 0, \quad \text{with} \quad \nabla_\mu T_{j}^{\mu\nu} = \frac{Q_j u_{\rm DM}^\nu}{a}.
    \label{eq:conservation}
\end{equation}

\noindent Here, the index $j$ represents DE or DM, $a$ denotes the scale factor, $u_{\rm DM}^\nu$ represents the four-velocity of DM, and $Q$ denotes the interaction rate between DE and DM.  
Starting from Eq.~\eqref{eq:conservation} and assuming a flat Friedmann–Lemaître–Robertson–Walker (FLRW) metric, we obtain the evolution equations for the energy densities of DM ($\rho_{\rm DM}$) and DE ($\rho_{\rm DE}$) in IDE models, given by
\begin{eqnarray}
\frac{d\rho_{\rm DM}}{d\tau} + 3 \mathcal{H} \rho_{\rm DM} &=& Q, \\
\frac{d\rho_{\rm DE}}{d\tau} + 3 \mathcal{H} (1 + w_{\rm DE}) \rho_{\rm DE} &=& -Q.
\end{eqnarray}
In these expressions, $\mathcal{H}$ represents the conformal Hubble parameter, $w_{\rm DE}$ denotes the equation-of-state parameter for DE, and $Q$ corresponds to the same interaction term introduced in Eq.~\eqref{eq:conservation}. The choice of $Q$ is purely phenomenological, given our limited knowledge of the dark sector.\footnote{Several distinct parametrization proposals exist in the literature; see, for example,~\cite{Gavela:2010tm,DiValentino:2017iww,Zhai_2023,Caprini_2016,Liu_2022,Sharov_2016,Feng_2016,Aljaf_2021,Zhai:2025hfi, Paliathanasis:2024abl, vanderWesthuizen:2025iam}.}

The introduction of a coupling between DM and DE also affects the evolution of perturbations. In the Newtonian gauge, the coupled system of linear Einstein-Boltzmann equations governing the evolution of DM density perturbations, $\delta_c$, and velocity divergences, $\theta_c$, is well established in the literature (see, e.g.,~\cite{Valiviita:2008iv, LopezHonorez:2012fq, Gavela:2010tm}). In the mildly nonlinear regime, modifications to the DM continuity equation must be taken into account, since the coupling alters the underlying mass-conservation dynamics and thereby impacts structure formation through changes in both the background expansion and the linear growth rate—two key ingredients in modeling the matter power spectrum. 

On the other hand, as shown by \cite{Dakin:2019vnj, Hassani:2020buk}, DE perturbations may, in principle, induce corrections to the Euler equation in interacting dark-sector scenarios; however, such effects become relevant only on extremely large scales, where a fully relativistic treatment is required. DE perturbations are dynamically relevant only on scales comparable to the cosmological horizon. On subhorizon scales, the large effective sound speed ($c_s^2 \sim 1$) generates a strong pressure that suppresses the growth of $\delta_{DE}$, rendering its contribution to the gravitational potential negligible when compared to that of DM. As a consequence, in large-scale structure analyses, DE perturbations can be consistently neglected. As a proof of concept in the context of IDE models, see Fig.~1 of Ref.~\cite{Silva:2024ift}, where the contribution of DE perturbations to the matter power spectrum is below the $1\%$ level for the range of scales relevant to this work. Because our analysis is confined to the Newtonian regime relevant for LSS observables, these contributions are safely negligible and will not be included here.

In general, by considering the Newtonian limit for nonrelativistic fluids—such as cold DM, while performing the appropriate coordinate transformations in an expanding universe, we obtain the standard continuity equation \cite{1980lssu.book.....P}
\begin{equation}
    \rho' + 3\mathcal{H}\rho + \nabla \cdot (\rho \mathbf{v}) = 0,
    \label{eq:continuidade_expansao}
\end{equation}
where a prime denotes derivative with respect to conformal time $\tau$, and $\rho$ and $\mathbf{v}$ denote the density and peculiar velocity of the fluid, respectively.  

When a fluid component $i$ interacts with other sectors (e.g., energy transfer from DE to DM), energy conservation is no longer preserved. In this case, we introduce a interaction term $Q(\tau, \mathbf{x})$, which represents the energy transfer rate per unit of physical volume
\begin{equation}
    \rho'_i + 3\mathcal{H}\rho_i +  \nabla \cdot (\rho_i \mathbf{v}_i) = Q(\tau, \mathbf{x}).
    \label{eq:continuidade_interacao}
\end{equation}

Building upon this framework, we introduce small perturbations in the matter density field and in the interaction source term, such that
\begin{equation}
\rho_i(\mathbf{x}, \tau) = \rho_i(\tau)\,[1 + \delta_i(\mathbf{x}, \tau)],
\qquad
Q(\mathbf{x}, \tau) = Q(\tau) + \delta Q(\mathbf{x}, \tau).
\end{equation}
Substituting these expressions into Eq.~\eqref{eq:continuidade_interacao}, we obtain the perturbed continuity equation for the interacting fluid ${\rm CDM}$—in this case, DM
\begin{equation}
\frac{\partial \delta_{\rm CDM}(\mathbf{x}, \tau)}{\partial \tau}
+ \nabla \cdot \Big\{ \big[1 + \delta_{\rm CDM}(\mathbf{x}, \tau)\big]\, \mathbf{v}_{\rm CDM}(\mathbf{x}, \tau) \Big\}
= \frac{\delta Q(\mathbf{x}, \tau)}{\rho_{\rm CDM}(\tau)}
- \frac{Q(\tau)}{\rho_c(\tau)}\, \delta_{\rm CDM}(\mathbf{x}, \tau),
\label{eq:perturbacao_nao_linear}
\end{equation}
Equation~\eqref{eq:perturbacao_nao_linear} explicitly reveals the modification of the nonlinear contributions that arise when considering the presence of a coupling between DE and DM. In what follows, we will simply omit the subscript “\text{CDM}”, as it is standard practice and implicitly understood that the fluid dynamic equations discussed below refer to cold DM.

It is important to emphasize that, within the class of IDE models of primary interest in this work, the Euler equation remains formally unchanged with respect to the $\Lambda$CDM framework. For instance, even in a fully relativistic treatment, it has been shown that the DM velocity-perturbation equation coincides with that of the uncoupled case (see Eqs.~(50) and (95) of Ref.~\cite{Valiviita:2008iv}). Similar conclusions are reached in other perturbative analyses (see, e.g., Ref.~\cite{Honorez_2010}). This behavior follows from the common assumption that the interaction four-vector satisfies $Q^\nu \propto u^\nu$, typically aligned with the four-velocity of either DM or DE. Under this assumption, no momentum transfer orthogonal to the fluid four-velocity is introduced, and hence no additional effective force appears in the Euler equation. As a result, DM continues to follow geodesic trajectories, the emergence of a fifth force is avoided, and the stability of cosmological perturbations is preserved. Therefore, the absence of modifications to the Euler equation in many IDE models is not a generic consequence of the interaction itself, but rather the outcome of a specific physical hypothesis regarding the nature of the coupling. On the other hand, an alternative approach explored in the literature considers a pure momentum exchange between the dark components (see, e.g., Ref.~\cite{Kumar:2017bpv, BeltranJimenez:2025yad, BeltranJimenez:2021wbq, Lague:2024sox}). In this scenario, the opposite situation is postulated: the full background dynamics and the continuity equations remain unchanged, while the interaction manifests itself only at the level of velocity perturbations, i.e., through modifications of the Euler equations.

In this work, we focus on the phenomenological class of IDE models characterized by an energy-exchange coupling $Q$ that does not modify the Euler equation, which has been the most widely studied scenario in the literature in recent years. Extensions to models involving pure momentum exchange between the dark components, and/or explicit modifications of the Euler equation, are certainly of interest and will be addressed in a future publication. Our primary objective here is to develop the one-loop corrections for some of the most commonly studied IDE classes in the literature. Consequently, within this framework, the Euler equation retains its standard form and can be written as

\begin{equation}
\frac{\partial \mathbf{v}(\mathbf{x}, \tau)}{\partial \tau}
+ \mathcal{H}(\tau)\, \mathbf{v}(\mathbf{x}, \tau)
+ \big[\mathbf{v}(\mathbf{x}, \tau) \cdot \nabla\big] \mathbf{v}(\mathbf{x}, \tau)
= -\nabla \Phi(\mathbf{x}, \tau),
\label{eq:euler}
\end{equation}
where $\Phi(\mathbf{x}, \tau)$ represents the gravitational potential.  
We also define here the velocity divergence, given by $\theta(\mathbf{x}, \tau) \equiv \nabla \cdot \mathbf{v}(\mathbf{x}, \tau)$, which will be useful in what follows.

The gravitational potential introduced in Eq.~\eqref{eq:euler} is determined by the Poisson equation, which retains its standard form within the IDE framework. This is because the underlying gravitational dynamics remain governed by general relativity, even in the presence of a coupling between DM and DE. Accordingly, we have  

\begin{equation}
\label{eq_poisson}
\nabla^2 \Phi(\mathbf{x}, \tau)
= \frac{3}{2}\, \mathcal{H}^2(\tau)\, \Omega_m(\tau)\, \delta(\mathbf{x}, \tau),
\end{equation}
where $\Omega_m(\tau)$ denotes the matter density parameter at conformal time $\tau$, and $\delta(\mathbf{x}, \tau)$ represents the total matter overdensity.

In the linear regime, $\delta(\mathbf{x}, \tau)$ appearing in Eq.~\eqref{eq_poisson} can be expressed in terms of the linear growth function, $D(\tau)$, which characterizes the temporal evolution of density perturbations driven by gravitational instability. It is defined as

\begin{equation}
\delta(\mathbf{x}, \tau) = D(\tau)\, \delta(\mathbf{x}, \tau_0)
\quad \therefore \quad
D(\tau) = \frac{\delta(\mathbf{x}, \tau)}{\delta(\mathbf{x}, \tau_0)},
\label{eq:gf}
\end{equation}
\noindent
where $\delta(\mathbf{x}, \tau_0)$ denotes the total matter density contrast at the conformal time corresponding to the present time. The growth function is conventionally normalized such that $D(z=0) = 1$.

Following the standard nonrelativistic perturbation theory approach, we combine Eq.~\eqref{eq:perturbacao_nao_linear} with Eqs.~\eqref{eq:euler} and \eqref{eq_poisson} to derive the linear growth equation in the presence of an interaction between DM and DE. To this end, we linearize the continuity and Euler equations, which contain nonlinear terms, while the Poisson equation \eqref{eq_poisson} is already linear in the density contrast. Working in conformal time and using the velocity divergence $\theta(\mathbf{x},\tau)$ defined above, the linearized continuity and Euler equations read as follows:
\begin{align}
\delta'(\mathbf{x},\tau) + \theta(\mathbf{x},\tau) &= S(\mathbf{x}, \tau), \label{lin_cont}\\
\theta'(\mathbf{x},\tau) + \mathcal{H}(\tau)\,\theta(\mathbf{x},\tau) &= -\nabla^2\Phi(\mathbf{x},\tau). \label{lin_euler}
\end{align}
The source term $S(\tau,\mathbf{x})$, which encapsulates the effect of the dark-sector coupling, is given by
\begin{equation}
S(\mathbf{x}, \tau) \equiv 
\frac{\delta Q(\mathbf{x}, \tau)}{\rho(\tau)}
- \frac{Q(\tau)}{\rho(\tau)}\,\delta(\mathbf{x},\tau).
\end{equation}

Using Poisson's equation and eliminating $\theta$ between Eqs.~\eqref{lin_cont} and \eqref{lin_euler}, we obtain the linear second-order equation for the dark-matter density contrast
\begin{equation}
\delta''(\mathbf{x},\tau)
+ \mathcal{H}(\tau)\,\delta'(\mathbf{x},\tau)
- \tfrac{3}{2}\,\mathcal{H}^2(\tau)\,\Omega_m(\tau)\,\delta(\mathbf{x},\tau)
= S'(\mathbf{x}, \tau) + \mathcal{H}(\tau)\,S(\mathbf{x}, \tau)\,,
\label{delta_linear_source}
\end{equation}
Building upon the definition in Eq.~\eqref{eq:gf}, we have that the linear growth function $D(\tau)$ in the presence of a dark-sector coupling satisfies the inhomogeneous second-order differential equation
\begin{equation}
\begin{split}
D''(\tau) + \mathcal{H}(\tau)\,D'(\tau) - \tfrac{3}{2}\,\mathcal{H}^2(\tau)\,\Omega_m(\tau)\,D(\tau)
= \frac{1}{\delta(\tau_0)} \Bigg[ 
\left(\frac{\delta Q(\tau)}{\rho(\tau)} - \frac{Q(\tau)}{\rho(\tau)}\,\delta(\tau)\right)' + \mathcal{H}(\tau)\left(\frac{\delta Q(\tau)}{\rho(\tau)} - \frac{Q(\tau)}{\rho(\tau)}\,\delta(\tau)\right) \Bigg],
\end{split}
\label{D_equation}
\end{equation}
where the right-hand side acts as a source term generated by perturbations in the interaction rate. In the limit $Q\to 0$ (and consequently $\delta Q\to 0$), Eq.~\eqref{D_equation} reduces to the standard homogeneous growth equation of $\Lambda$CDM model.

The results reviewed in this section illustrate how IDE can modify both the background dynamics and the evolution of perturbation modes in the linear regime.

\section{Nonlinear Perturbation Theory in IDE Cosmology}
\label{sec:nl_IDE}

In this section, we compute the one-loop corrections to the density contrast $\delta$, following the standard SPT procedure~\cite{Bernardeau:2001qr}. Defining the physical divergence of the peculiar velocity as
\(\theta \equiv \nabla \cdot \mathbf{v}\), we can rewrite respectively equations (\ref{lin_cont}) and (\ref{lin_euler}) in the nonlinear regime

\begin{equation}\label{cont:nlinear}
\frac{\partial \delta (\mathbf{x},\tau)}{\partial \tau} + \theta(\mathbf{x},\tau) 
= - \nabla \cdot (\delta \mathbf{v}) 
+ \frac{\delta Q(\mathbf{x},\tau)}{\rho(\tau)} - \frac{Q(\tau)}{\rho(\tau)}\delta(\mathbf{x},\tau),
\end{equation}
\begin{equation}
\frac{\partial \theta(\mathbf{x},\tau)}{\partial \tau} 
+ \mathcal{H}(\tau) \theta(\mathbf{x},\tau) + \nabla^2 \Phi(\mathbf{x},\tau) 
= \nabla \cdot \big[(\mathbf{v} \cdot \nabla) \mathbf{v}\big].
\label{eq:realspace_euler}
\end{equation}

In this work, we adopt the following convention for the Fourier transform
\[
f(\mathbf{x}) = \int \frac{d^3k}{(2\pi)^3}\, e^{i \mathbf{k}\cdot \mathbf{x}}\, f(\mathbf{k}),
\qquad
f(\mathbf{k}) = \int d^3x \, e^{-i \mathbf{k}\cdot \mathbf{x}}\, f(\mathbf{x}).
\]

Taking the Fourier transform of Eq.~(\ref{cont:nlinear}), we obtain
\begin{equation}
\delta'(\mathbf{k},\tau) + \theta(\mathbf{k},\tau) = \int \frac{d^{3}q_{1}}{(2\pi)^{3}} \frac{d^{3}q_{2}}{(2\pi)^{3}} (2\pi)^{3}\,\alpha(\mathbf{q}_{1},\mathbf{q}_{2})\,\delta_{D}(\mathbf{k}-\mathbf{q}_{1}-\mathbf{q}_{2})\,\theta(\mathbf{q}_{1},\tau)\,\delta(\mathbf{q}_{2},\tau) + \left[ \frac{\delta Q(\mathbf{k},\tau)}{\rho(\tau)} - \frac{Q(\tau)}{\rho(\tau)}\,\delta(\mathbf{k},\tau) \right],
\label{eq:fourier-irrot}
\end{equation}
where 
\begin{equation}
\alpha(\mathbf{q}_1, \mathbf{q}_2) = 1 + \frac{\mathbf{q}_1 \cdot \mathbf{q}_2}{q_1^2},
\end{equation}
is precisely the standard form of the $\alpha$ mode coupling that arises naturally from the nonlinear terms in the continuity equation. Physically, this coupling reflects the fact that, beyond the linear regime, modes with different wave vectors $\mathbf{q}_1$ and $\mathbf{q}_2$ no longer evolve independently but instead interact and exchange power. In the above equation, $\delta Q(\mathbf{k}, \tau)$ represents the Fourier transform of $\delta Q(\mathbf{x}, \tau)$.

The Eq. (\ref{eq:fourier-irrot}) generalizes the standard SPT continuity equation to interacting DE models, capturing both gravitational nonlinearities and energy exchange effects in Fourier space. It is worth reiterating that the IDE framework considered here does not introduce any modifications to the Euler equation. As a result, the nonlinear equation associated with the velocity field remains unchanged from its standard form, and his representation in the Fourier space is
\begin{equation}
\frac{\partial \theta(\mathbf{k},\tau)}{\partial \tau} + \mathcal{H}(\tau)\,\theta(\mathbf{k},\tau) + \frac{3}{2}\,\Omega_m(\tau)\,\mathcal{H}^2(\tau)\,\delta(\mathbf{k},\tau) = - \int \frac{d^{3}q_{1}}{(2\pi)^{3}} \frac{d^{3}q_{2}}{(2\pi)^{3}} (2\pi)^{3}\,\delta_{D}(\mathbf{k}-\mathbf{q}_{1}-\mathbf{q}_{2})\,\beta(\mathbf{q}_{1},\mathbf{q}_{2})\,\theta(\mathbf{q}_{1},\tau)\,\theta(\mathbf{q}_{2},\tau),
\label{Eq_21}
\end{equation}
where
\begin{equation}
    \beta(\mathbf{q}_1, \mathbf{q}_2) = \frac{(\mathbf{q}_1 \cdot \mathbf{q}_2)(q_1^2 + q_2^2)}{2 q_1^2 q_2^2},
\end{equation}
is the other standard mode coupling $\beta$. 

By inspecting Eqs.~\eqref{eq:fourier-irrot} and~\eqref{Eq_21}, one immediately sees that any nonlinear modification within the IDE framework originates solely from the altered continuity equation. To proceed and assess its implications, we follow the standard approach in the literature and adopt a phenomenological parametrization for the coupling function $Q$.

At this point, we can define the extra source term in Eq.~(\ref{eq:fourier-irrot}) as
\begin{equation}
\label{S_main_eq}
S(\mathbf{k}, \tau) = \frac{\delta Q(\mathbf{k}, \tau)}{\rho(\tau)}
- \frac{Q(\tau)}{\rho(\tau)} \, \delta(\mathbf{k}, \tau).
\end{equation}

We start from a covariant expression for the interaction four-vector $Q^{\mu}$, allowing for a systematic identification of the background energy transfer rate $Q$ and its perturbation $\delta Q$. We consider the following ansatz:
\begin{equation}
    Q^{\mu}(\mathbf{x},\tau) = -\xi\,\mathcal{H}(\tau)\,\rho_{m}(\mathbf{x},\tau)\,u_c^{\mu}(\mathbf{x},\tau)
\end{equation}
where \( u_c^{\mu} = a^{-1}(1-\Phi,\,\partial^i v) \) is the four-velocity of the CDM fluid, \( v \) denotes the peculiar velocity potential, \( \Phi\) is the gravitational potential, and \( \xi \) is a dimensionless constant that parametrizes the strength of the dark-sector interaction.

The crucial step is to compute the temporal component of $Q^{\mu}$ and confront it with the general decomposition of the energy--momentum transfer four-vector (see \cite{Valiviita:2008iv} for a detailed discussion). By matching both expressions, we can directly identify the background and perturbed energy transfer rates as
\begin{align}
    Q(\tau) &= \xi\,\mathcal{H}\,\rho_m(\tau), \\
    \delta Q(\mathbf{x},\tau) &= \xi\,\mathcal{H}\,\rho_m(\tau)\,\delta(\mathbf{x},\tau).
\end{align}

The background evolution equation for DM is then given by
\begin{equation}
\dot{\rho}_m + (3 - \xi) \mathcal{H} \rho_m = 0,
\label{eq:background_interaction}
\end{equation}
whose solution is
\begin{equation}
\rho_m \propto a^{-3 + \xi},
\end{equation}
this result shows that DM decays: more slowly than \(a^{-3}\) if \(\xi > 0\) (energy gain), and more rapidly than \(a^{-3}\) if \(\xi < 0\) (energy loss). 

\begin{comment}
As a first example, we consider the well-known interaction term of the form:
\begin{equation}
Q = \xi \mathcal{H} \rho_m,
\end{equation}
where \(\xi\) is a constant parameter measuring the strength of the interaction. The background equation for DM is then
\begin{equation}
\dot{\rho}_m + (3 - \xi) \mathcal{H} \rho_m = 0,
\label{eq:background_interaction}
\end{equation}
whose solution is
\begin{equation}
\rho_m \propto a^{-3 + \xi},
\end{equation}
this result shows that DM decays: more slowly than \(a^{-3}\) if \(\xi > 0\) (energy gain), and more rapidly than \(a^{-3}\) if \(\xi < 0\) (energy loss).  

For the perturbation of the interaction term, we have  
\begin{equation}
\delta Q = \xi \mathcal{H} \rho_m \delta_m.
\end{equation}
\end{comment}
Substituting the expressions for \( Q \) and \( \delta Q \) in Eq. (\ref{S_main_eq}), we then find that  
\begin{equation}
S(\mathbf{k}, \tau) = 0,
\end{equation}
and thus we recover the continuity equations in the nonlinear regime for the standard \(\Lambda\)CDM model.  
\textit{This result highlights an important feature: certain forms of interaction do not affect the growth of structures in the nonlinear regime.}

If the interaction were instead of the form \(Q = \xi \mathcal{H} \rho_X\) (with a different component \(X\)), we would have
\begin{equation}
S(\mathbf{k}, \tau) = \xi \mathcal{H} \frac{\rho_X}{\rho_m} \, (\delta_X - \delta_m),
\end{equation}
which introduces a coupling between the perturbations of the two components. Naturally, for the IDE scenarios, we will have $X = DE$. The final solution of the continuity equation, including the interaction term \(Q = \xi \mathcal{H}\rho_{\rm DE}\), is
\begin{equation}
\begin{split}
\delta'_m(\mathbf{k},\tau) + \theta_m(\mathbf{k},\tau) 
&= 
\int \frac{d^3q_1}{(2\pi)^3} \frac{d^3q_2}{(2\pi)^3} \,
(2\pi)^3 \delta_{\rm D}(\mathbf{k}-\mathbf{q}_1-\mathbf{q}_2) \,
\alpha(\mathbf{q}_1,\mathbf{q}_2) \,
\delta_m(\mathbf{q}_1, \tau) \, \theta_m(\mathbf{q}_2, \tau) \\
&\quad + \, \left(
\frac{\xi \mathcal{H} \rho_{\rm DE} \, \delta_{\rm DE}(\mathbf{k},\tau)}{\rho_m} 
- \frac{\xi \mathcal{H} \rho_{\rm DE}}{\rho_m} \, \delta_m(\mathbf{k},,\tau)
\right).
\end{split}
\end{equation}

Since as we discussed, is expected that DE only cluster on extremely high scales, in that case $\delta_{\rm DE} = 0$. So our final continuity equation take the form
\begin{equation}
\label{delta_f_nl}
\delta'_m(\mathbf{k},\tau) + \theta_m(\mathbf{k},\tau) 
= \int \frac{d^3 q_1}{(2\pi)^3} \frac{d^3 q_2}{(2\pi)^3} \,
(2\pi)^3 \delta_{\rm D}(\mathbf{k}-\mathbf{q}_1-\mathbf{q}_2) \,
\alpha(\mathbf{q}_1, \mathbf{q}_2) \, 
\delta_m(\mathbf{q}_1, \tau) \theta_m(\mathbf{q}_2, \tau) 
+ S_m(\mathbf{k},\tau),
\end{equation}
where the linear source term due to DM–DE coupling is given by
\begin{equation}
S_m(\mathbf{k},\tau) = - \, \frac{\xi \mathcal{H} \rho_{\rm DE}}{\rho_m}  \delta_m(\mathbf{k},\tau).
\end{equation}

In this form, the nonlinear convolution term retains the standard advective interaction via the mode coupling \(\alpha\), while the coupling to the IDE sector appears explicitly as an additional linear source in the evolution equations. 

In what follows, for the subsequent developments, we need to compute the coupling kernel between the wave numbers. Following the standard procedure (see Ref.~\cite{Bernardeau:2001qr}), which consists of solving the perturbative equations (\ref{eq:fourier-irrot}) and (\ref{Eq_21}), we find that, in this specific case, the IDE effect is not merely associated with a new effective mode coupling. In other words, the additional contribution from $S_m(\mathbf{k},\tau)$ cannot be simply absorbed into the standard mode-coupling $\alpha$ and/or $\beta$. After performing extensive algebraic manipulations, we find that the recursive relations for the kernels $F_n$ and $G_n$ deviate from their standard forms and are given by

\begin{equation}
\begin{split}
F_n(\mathbf{q}_1,\ldots,\mathbf{q}_n,\tau)
&= \sum_{m=1}^{n-1} 
\frac{G_m(\mathbf{q}_1,\ldots,\mathbf{q}_m)}
     {(2n+3)(n-1) + \lambda(2n+1)}
\\[4pt]
&\quad \times 
\left[ (2n+1)\,\alpha(\mathbf{k}_1,\mathbf{k}_2)\,
       F_{\,n-m}(\mathbf{q}_{m+1},\ldots,\mathbf{q}_n)
       + 2\,\beta(\mathbf{k}_1,\mathbf{k}_2)\,
       G_{\,n-m}(\mathbf{q}_{m+1},\ldots,\mathbf{q}_n)
\right],
\label{Fn}
\end{split}
\end{equation}\\

\begin{equation}
\begin{split}
G_n(\mathbf{q}_1,\ldots,\mathbf{q}_n, \tau)
&= \sum_{m=1}^{n-1} 
\frac{G_m(\mathbf{q}_1,\ldots,\mathbf{q}_m)}
     {(2n+3)(n-1) + \lambda(2n+1)}
\\[4pt]
&\quad \times
\left[ 3\,\alpha(\mathbf{k}_1,\mathbf{k}_2)\,
       F_{\,n-m}(\mathbf{q}_{m+1},\ldots,\mathbf{q}_n)
       + (n+\lambda)\,2\,\beta(\mathbf{k}_1,\mathbf{k}_2)\,
       G_{\,n-m}(\mathbf{q}_{m+1},\ldots,\mathbf{q}_n)
\right],
\label{Gn}
\end{split}
\end{equation}\\

\noindent where we have defined $\lambda \equiv \frac{\xi \rho_{\mathrm{de}}}{\rho_m}$, which fully captures the deviation from the standard model. For $\xi = 0$, we recover the default case, as expected. We also have $\mathbf{k_1 = q_1 + \ldots + q_m}$, $\mathbf{k_2 = q_{m+1} + \ldots + q_n}$, $\mathbf{k = k_1 + k_2}$, $\mathbf{q_{ij} = q_i + q_j}$, and $F_1 = G_1 = 1$. 

In Appendix A, we explicitly present the form of the one-loop correction under this parametrization.

On the other hand, nonlinear interaction terms in the coupling function $Q$ are also commonly studied in the literature. In this context, such terms are particularly well suited for probing the impact of a nonlinear perturbative treatment in IDE scenarios. Motivated by this, we introduce a phenomenological nonlinear interaction term of the form:
\begin{equation}
Q(\mathbf{x}, \tau) = \Gamma \, \rho_m(\tau) \, \rho_{\rm DE}(\tau) \, \theta_m(\mathbf{x}, \tau),
\label{eq:paramet_new}
\end{equation}
where \(\Gamma\) is the nonlinear coupling constant in the dark sector (with units of volume per energy), and \(\theta_m = \nabla \cdot \mathbf{v}_m\) denotes the divergence of the DM peculiar velocity. 

Starting from this phenomenological parametrization, we construct a covariant generalization of the interaction, following the same procedure outlined above. We adopt the ansatz
\begin{equation}
    Q^{\mu} = \Gamma\,\rho_m(\mathbf{x},\tau)\,\rho_{\mathrm{DE}}(\mathbf{x},\tau)\,u^{\mu}\,\theta(\mathbf{x},\tau),
\end{equation}
where $u^{\mu}$ denotes the four-velocity of the matter fluid and $\theta$ is a scalar interaction field.

By construction, the velocity field $\theta$ has no homogeneous component, so that the interaction vanishes at the background level, $Q=0$. Therefore, the energy transfer arises purely from perturbations and is intrinsically nonlinear.

Neglecting DE perturbations, the leading contribution to the interaction is given by
\begin{equation}
\delta Q = \Gamma\,\rho_m\,\rho_{\rm DE}\,\delta(\mathbf{x},\tau)\,\theta(\mathbf{x},\tau),
\end{equation}
which is explicitly second order in perturbations.

Therefore, we are left only with a purely nonlinear source component:

\begin{equation}
S_m^{\rm NL} = \Gamma \, \rho_{\rm DE} \, \delta_m \, \theta_m.
\label{Sm}
\end{equation}

The Fourier transform of the nonlinear source term, Eq.~(\ref{Sm}), can then be written as
\begin{equation}
S_m^{\rm NL} = \int \frac{d^3q_1}{(2\pi)^3} \frac{d^3q_2}{(2\pi)^3} (2\pi)^3 \delta_D(\mathbf{k}-\mathbf{q}_1-\mathbf{q}_2) 
\alpha_{\rm IDE}(\tau) \, \delta_m(\mathbf{q}_1, \tau) \, \theta_m(\mathbf{q}_2, \tau),
\end{equation}
where we have defined
\begin{equation}
\alpha_{\rm IDE}(\tau) = \Gamma \, \rho_{\rm DE}(\tau).
\end{equation}

Therefore, the modified continuity equation, Eq.~(\ref{delta_f_nl}), can be recast in terms of a single effective mode-coupling,
\begin{equation}
\alpha_{\rm eff}(\mathbf{q}_{1},\mathbf{q}_{2},\tau)
    \equiv \alpha(\mathbf{q}_{1},\mathbf{q}_{2})
    + \alpha_{\rm IDE}(\tau)\,,
\end{equation}
where $\alpha$ denotes the standard SPT mode-coupling and $\alpha_{\rm IDE}$ encapsulates the IDE-induced correction arising from the nonlinear interaction term.

Using this definition, the full continuity equation becomes
\begin{equation}
\label{delta_IDE_f}
\delta'_m(\mathbf{k},\tau) + \theta_m(\mathbf{k},\tau)
= \int \frac{d^{3}q_{1}}{(2\pi)^{3}} \frac{d^{3}q_{2}}{(2\pi)^{3}}
(2\pi)^{3} \delta_{D}(\mathbf{k}-\mathbf{q}_{1}-\mathbf{q}_{2})\,
\alpha_{\rm eff}(\mathbf{q}_{1},\mathbf{q}_{2},\tau)\,
\delta_m(\mathbf{q}_{1},\tau)\,\theta_m(\mathbf{q}_{2},\tau).
\end{equation}

In this representation, the IDE model preserves the standard SPT structure while modifying only the strength of the mode coupling through $\alpha_{\rm eff}$. As expected, the usual $\Lambda$CDM expression is recovered when $\Gamma = 0$, for which $\alpha_{\rm eff} = \alpha$.

\section{One-Loop Corrections to the Power Spectrum}
\label{sec:1L_IDE}
In this section, we investigate how the modifications to the matter density contrast propagate into the matter power spectrum within the IDE framework. In particular, we assess how the interaction parametrization introduced in Eq.~\eqref{eq:paramet_new} alters the evolution of perturbations and, consequently, the resulting shape of $P(k)$ when compared to the standard noninteracting case.

Following the standard procedure \cite{Bernardeau:2001qr}, the density contrast can be expanded perturbatively as

\begin{equation}
\delta_n(\mathbf{k}) = \int \frac{d^3q_1}{(2\pi)^3} \cdots \frac{d^3q_n}{(2\pi)^3} (2\pi)^3\, \delta_D\Bigg( \mathbf{k} - \sum_{i=1}^n \mathbf{q}_i \Bigg) 
F_n(\mathbf{q}_1, \ldots, \mathbf{q}_n) \, \delta_1(\mathbf{q}_1) \cdots \delta_1(\mathbf{q}_n),
\end{equation}
where \(F_n\) are the standard perturbative kernels at order \(n\), and \(\delta_1(\mathbf{q})\) is the linear density contrast in Fourier space.

In particular, we are interested in computing the contributions to the matter power spectrum arising from the second- and third-order terms, \(\delta_2(\mathbf{k})\) and \(\delta_3(\mathbf{k})\), whose Fourier-space expressions are

\begin{equation}
\delta_2(\mathbf{k}) = \int \frac{d^3q_1}{(2\pi)^3} \frac{d^3q_2}{(2\pi)^3} \, (2\pi)^3 \delta_D(\mathbf{k} - \mathbf{q}_1 - \mathbf{q}_2) 
F_2(\mathbf{q}_1, \mathbf{q}_2) \, \delta_1(\mathbf{q}_1) \, \delta_1(\mathbf{q}_2),
\end{equation}
and
\begin{equation}
\delta_3(\mathbf{k}) = \int \frac{d^3q_1}{(2\pi)^3} \, \frac{d^3q_2}{(2\pi)^3} \, \frac{d^3q_3}{(2\pi)^3} (2\pi)^3 \, \delta_D(\mathbf{k} - \mathbf{q}_1 - \mathbf{q}_2 - \mathbf{q}_3) 
F_3(\mathbf{q}_1, \mathbf{q}_2, \mathbf{q}_3) \, \delta_1(\mathbf{q}_1) \, \delta_1(\mathbf{q}_2) \, \delta_1(\mathbf{q}_3).
\end{equation}

At this point we can define the correlations functions that appears on the 1-loop power spectrum expansion as
\begin{equation}
\langle \delta_2(\mathbf{k}) \delta_2(\mathbf{k}') \rangle = (2\pi)^3 \delta_D(\mathbf{k} + \mathbf{k}') P_{22}(k), \qquad \langle \delta_{1}(\mathbf{k}) \delta_3(\mathbf{k}') \rangle = (2\pi)^3 \delta_D(\mathbf{k} + \mathbf{k}') P_{13}(k).
\end{equation}

For the first one-loop contribution, $P_{22}(k)$, performing the required algebraic manipulations yields
\begin{equation}\label{P22new}
P_{22}^{\rm IDE}(k) = 
2 \int d^3 q \,
\left[ F_2^{\rm IDE}(\mathbf{q}, \mathbf{k}-\mathbf{q}) \right]^2
P_L(|\mathbf{k}-\mathbf{q}|)\, P_L(q),
\end{equation}
where the effective second-order kernel is defined as
\begin{equation}
\label{F2_new}
F_2^{\rm IDE}(\mathbf{q}_1, \mathbf{q}_2)
= F_2(\mathbf{q}_1, \mathbf{q}_2)
+ \frac{5}{7}\, \alpha_{\rm IDE}(\tau),
\end{equation}
with $F_2$ denoting the standard second-order kernel and $\alpha_{\rm IDE}(\tau)$ encoding the nonlinear contribution associated with the IDE source term.

The expression above makes explicit that the kernel $F_2$ receives a nontrivial correction in the presence of IDE, rather than appearing as a simple rescaling of the standard kernel. Consequently, the computation of the IDE power spectrum requires replacing the kernels in Eq.~\eqref{F2_new} and subsequently inserting them into the definition of $P_{22}$, Eq.~\eqref{P22new}. Carrying out this substitution, the modified one-loop contribution becomes
\begin{equation}
\label{P22final}
P_{22}^{\rm IDE}(k) =
P_{22}(k)
+ \frac{20}{7}\, \alpha_{\rm IDE}(\tau)
\int d^3 q \,
F_2(\mathbf{q}, \mathbf{k}-\mathbf{q})\,
P_L(|\mathbf{k}-\mathbf{q}|)\, P_L(q),
\end{equation}
where $P_{22}(k)$ denotes the standard (noninteracting) contribution. As expected, in the limit $\Gamma = 0$ -- or equivalently $\alpha_{\rm IDE}=0$ -- the standard result is fully recovered.

A similar procedure can be applied to the \(P_{13}(k)\) contribution. The modified term is given by
\begin{equation}
\label{F_3_new}
\begin{aligned}
F_3^{\rm IDE}(\mathbf{q_1,q_2,q_3})
&= F_3(\mathbf{q_1,q_2,q_3})
\\[4pt]
&\quad
+ \frac{\alpha_{\rm IDE}(\tau)}{18} \Big[
    5\,\alpha(\mathbf{q_1,q_{23}})
    + 7\,F_2(\mathbf{q_2,q_3})
    + \tfrac{6}{7}\,\beta(\mathbf{q_1,q_{23}})
\\[2pt]
&\qquad\qquad\qquad\qquad
    + 3\,\alpha(\mathbf{q_1,q_2})
    + 4\,\beta(\mathbf{q_1,q_2})
\Big]
\\[6pt]
&\quad
+ \frac{\alpha_{\rm IDE}(\tau)}{18} \Big[
    \tfrac{6}{7}\,\beta(\mathbf{q_{12},q_3})
    + 3\,\alpha(\mathbf{q_{12},q_3})
\Big],
\end{aligned}
\end{equation}
with \(F_3\) being the standard third-order kernel.  

So, substituting Eq.~(\ref{F_3_new}) in the general expression, we obtain\footnote{We set  
\(\mathbf{q}_1 = \mathbf{k}\), \(\mathbf{q}_2 = \mathbf{q}\), \(\mathbf{q}_3 = -\mathbf{q}\),  
\(\mathbf{q}_{23} = \mathbf{q}_2 + \mathbf{q}_3 = \mathbf{0}\), and \(\mathbf{q}_{12} = \mathbf{q}_1 + \mathbf{q}_2 = \mathbf{k} + \mathbf{q}\).}

\begin{equation}
P_{13}^{\rm IDE}(k) = P_{13}(k) + \Delta P_{13}(k),
\end{equation}
where  
\begin{equation}
\label{new_Delta_P_13}
\Delta P_{13}(k)
= \frac{\alpha_{\rm IDE}(\tau)}{3}\,P_L(k)\int d^3q \;
\Big[\,5+3\,\alpha(\mathbf{k},\mathbf{q})
+ 4\,\beta(\mathbf{k},\mathbf{q}) + \frac{6}{7}\beta(\mathbf{k}+\mathbf{q},-\mathbf{q})
+ 3\,\alpha(\mathbf{k}+\mathbf{q},-\mathbf{q}) \Big]\,P_L(q).
\end{equation}

Here, we use the notation \(\Delta P_{13}(k)\) to quantify the total deviation from the standard case. We can then rewrite the integrand in Eq.~(\ref{new_Delta_P_13}) as

\begin{equation}
\mathcal{I}(k,q,\mu) \equiv
\Big[\,5+3\,\alpha(\mathbf{k},\mathbf{q})
+ 4\,\beta(\mathbf{k},\mathbf{q}) + \frac{6}{7}\beta(\mathbf{k}+\mathbf{q},-\mathbf{q})
+ 3\,\alpha(\mathbf{k}+\mathbf{q},-\mathbf{q}) \Big],
\end{equation}
which simplifies to  
\begin{equation}
\mathcal{I}(k,q,\mu) 
= 11 + \frac{3q}{k}\mu + 4\beta(k,q,\mu) +\frac{6}{7}\frac{(-kq\mu -q^2)(k^2 + 2q^2 + 2kq)}{2q^2(k + q)^2}
- 3\,\frac{q^2+kq\mu}{(k+q)^2},
\end{equation}
after orienting \(\mathbf{k}\) along the \(z\)-axis and introducing \(\mu=\hat{\mathbf{k}}\cdot\hat{\mathbf{q}}\).  

The full integral then becomes  
\begin{equation}
\int d^3q\;\mathcal{I}(k,q,\mu) P_L(q)
= 2\pi\int dq\,q^2 P_L(q)\int_{-1}^{1} d\mu\; \mathcal{I}(k,q,\mu).
\end{equation}

After performing the angular integration, one finds the approximation  

\begin{equation}
\label{delta_P13_final}
\Delta P_{13}(k)
= \frac{2\pi\alpha_{\rm int}(\tau)}{3}\,P_L(k)\int dq\,q^2\,P_L(q)\,
\left[\,11-\frac{3}{7}A(k,q)\,-3B(k,q)\,\right],
\end{equation}
where 

\begin{equation}
A(k,q) = \frac{k^2+2q^2+2kq}{(k+q)^2},
\end{equation}
and
\begin{equation}
B(k,q) =  \frac{q^2}{(k + q)^2}.
\end{equation}
Equation~(\ref{delta_P13_final}) provides a practical integral to evaluate the additional correction to \(P_{13}\) arising from the IDE coupling considered in this work. This method we present is very general, with no approximation, and can be used to compute integrals in any modified theory of gravity.

In summary, in the mildly nonlinear regime and in real space, the full one-loop matter power spectrum in the presence of IDE can be expressed as the sum of two contributions

\begin{equation}
\label{full_1_loop}
P_{\rm 1 - loop}(k)
= P_{\rm 1 - loop,IDE}(k) + P_{\rm ctr}(k),
\end{equation}
where
\begin{equation}
P_{\rm 1 - loop,IDE}(k)
= P_{22}(k) + P_{13}(k)
+ \Delta P_{22}(k) + \Delta P_{13}(k).
\end{equation}
The terms $\Delta P_{22}$ and $\Delta P_{13}$ encode the IDE–induced departures from the standard one-loop result. In particular,
\begin{equation}
\Delta P_{22}(k)
= \frac{20}{7}\,\alpha_{\rm IDE}(\tau)
\int d^3 q \;
F_2(\mathbf{q},\mathbf{k}-\mathbf{q})\,
P_L(|\mathbf{k}-\mathbf{q}|)\,P_L(q),
\end{equation}
while $\Delta P_{13}(k)$ corresponds to the approximation introduced in Eq.~\eqref{delta_P13_final}.

In Eq.~(\ref{full_1_loop}), the term \(P_{\mathrm{ctr}}(z, k)\) denotes the standard counterterm at this order in perturbation theory, accounting for small-scale effects captured by the EFTofLSS~\cite{Baumann:2010tm, Senatore:2014vja, Carrasco:2013mua}. 

Additionally, the divergent velocity kernels at second and third order can be written as
\begin{equation}
    G_2^{\rm IDE}(\mathbf{q_1,q_2}) = G_2(\mathbf{q_1,q_2}) + \frac{3}{7}\alpha_{IDE}(\tau)
    \label{eq:G2}
\end{equation}
and 
\begin{equation}
\begin{aligned}
G_3^{\rm IDE}(\mathbf{q_1,q_2,q_3})
&= G_3(\mathbf{q_1,q_2,q_3})
\\[4pt]
&\quad
+ \frac{\alpha_{\rm IDE}(\tau)}{18} \Big[
    \frac{15}{7}\,\alpha(\mathbf{q_1,q_{23}})
    + 3\,F_2(\mathbf{q_2,q_3})
    + \frac{18}{7}\,\beta(\mathbf{q_1,q_{23}})
\\[2pt]
&\qquad\qquad\qquad\qquad
    + \frac{9}{7}\,\alpha(\mathbf{q_1,q_2})
    + \frac{12}{7}\,\beta(\mathbf{q_1,q_2})
\Big]
\\[6pt]
&\quad
+ \frac{\alpha_{\rm IDE}(\tau)}{18} \Big[
    \frac{18}{7}\,\beta(\mathbf{q_{12},q_3})
    + \frac{9}{7}\,\alpha(\mathbf{q_{12},q_3})
\Big].
\end{aligned}
\label{eq:G3}
\end{equation}
where $G_2$ and $G_3$ are the standard kernels. Again, for $\alpha_{\rm IDE}=0$, we recover the standard kernels. 

It is important to mention that, when transforming to redshift space, the general expression for the total one-loop corrections requires the fields $\delta$ and $\theta$, as well as their cross–correlations $\delta\theta$ and $\theta\theta$. In this case, the total one-loop power spectrum is given by

\begin{equation}
P_{\text{1-loop}}^{ab}(k) = P^{ab}_{22}(k) + P^{ab}_{13}(k),
\label{eq:1loop_Pab}
\end{equation}
where \(a\) and \(b\) denote either the \(\delta\) or \(\theta\) fields.

The calculations presented above correspond to the corrections associated with the density contrast $\delta$. The $\theta\theta$ power spectrum and the cross–correlation power spectrum $\delta\theta$ take the form

\begin{equation}
P_{22}^{\delta\theta}(k) = 2 \int d^3q \, 
F_2^{\rm IDE}(\mathbf{q}, \mathbf{k} - \mathbf{q}) \,
G_2^{\rm IDE}(\mathbf{q}, \mathbf{k} - \mathbf{q}) \,
P_L(q) \, P_L(|\mathbf{k} - \mathbf{q}|),
\end{equation}

\begin{equation}
P_{13}^{\delta\theta}(k) = 3 P_L(k) 
\int d^3q \, 
\left[
F_3^{\rm IDE}(\mathbf{k}, -\mathbf{q}, +\mathbf{q}) \,
+ G_3^{\rm IDE}(\mathbf{k}, -\mathbf{q}, +\mathbf{q})
\right] 
P_L(q),
\end{equation}

\begin{equation}
    P_{22}^{\theta \theta}(k) = 2 \int d^3q \, \left[ G_2^{\rm IDE}(\mathbf{q}, \mathbf{k} - \mathbf{q}) \right]^2 \, 
P_L(|\mathbf{k} - \mathbf{q}|) \, P_L(q),
\end{equation}

\begin{equation}
     P_{13}^{\theta \theta}(k) = 6P_L (k) \int d^3q \, G_3^{\rm IDE}(\mathbf{k}, \mathbf{-q}, \mathbf{q})  \, 
 \, P_L(q).
\end{equation}
which can be expanded and evaluated through Eqs.~\eqref{F2_new}, \eqref{F_3_new}, \eqref{eq:G2}, and \eqref{eq:G3}, following the same procedure applied to the $\delta$ fields. The full calculations are extremely lengthy, and for the sake of clarity, the final results are given by

\begin{equation}
\Delta P_{22} ^{\rm \theta \theta}(k) = -\frac{48  \pi}{49}\alpha_{\rm IDE}\int dq \ q^2 \frac{2q^2 + k^2-2kq}{(k-q)^2}P_L(| \mathbf{k-q}|)P_L (q)
\end{equation}

\begin{equation}\label{SEP}
\Delta P_{22} ^{\rm \delta \theta}(k) = \alpha_{\rm IDE}\left( \frac{6}{7}\int d^3q \, F_2(\mathbf{q}, \mathbf{k} - \mathbf{q}) \, P_L(|\mathbf{k} - \mathbf{q}|) \, P_L(q) -\frac{40\pi}{49}\int dq \ q^2 \frac{2q^2 + k^2-2kq}{(k-q)^2}P_L(| \mathbf{k-q}|)P_L (q) \right) ,
\end{equation}

\begin{equation}
\Delta P_{13} ^{\rm \theta \theta}(k) = \frac{2\pi\alpha_{\rm IDE}(\tau)}{3}\,P_L(k)\int dq\,q^2\,P_L(q)\,
\left[\frac{33}{7}- \frac{9}{7}A(k,q) - \frac{9}{7}B(k,q)\,\right],
\end{equation}

\begin{equation}
\Delta P_{13} ^{\rm \delta \theta}(k) = \frac{2\pi\alpha_{\rm IDE}(\tau)}{6}\,P_L(k)\int dq\,q^2\,P_L(q)\,
\left[\frac{110}{7}- \frac{12}{7}A(k,q) - \frac{30}{7}B(k,q)\,\right].
\end{equation}

\begin{figure}[htpb!]
    \centering
    \includegraphics[scale=0.36]{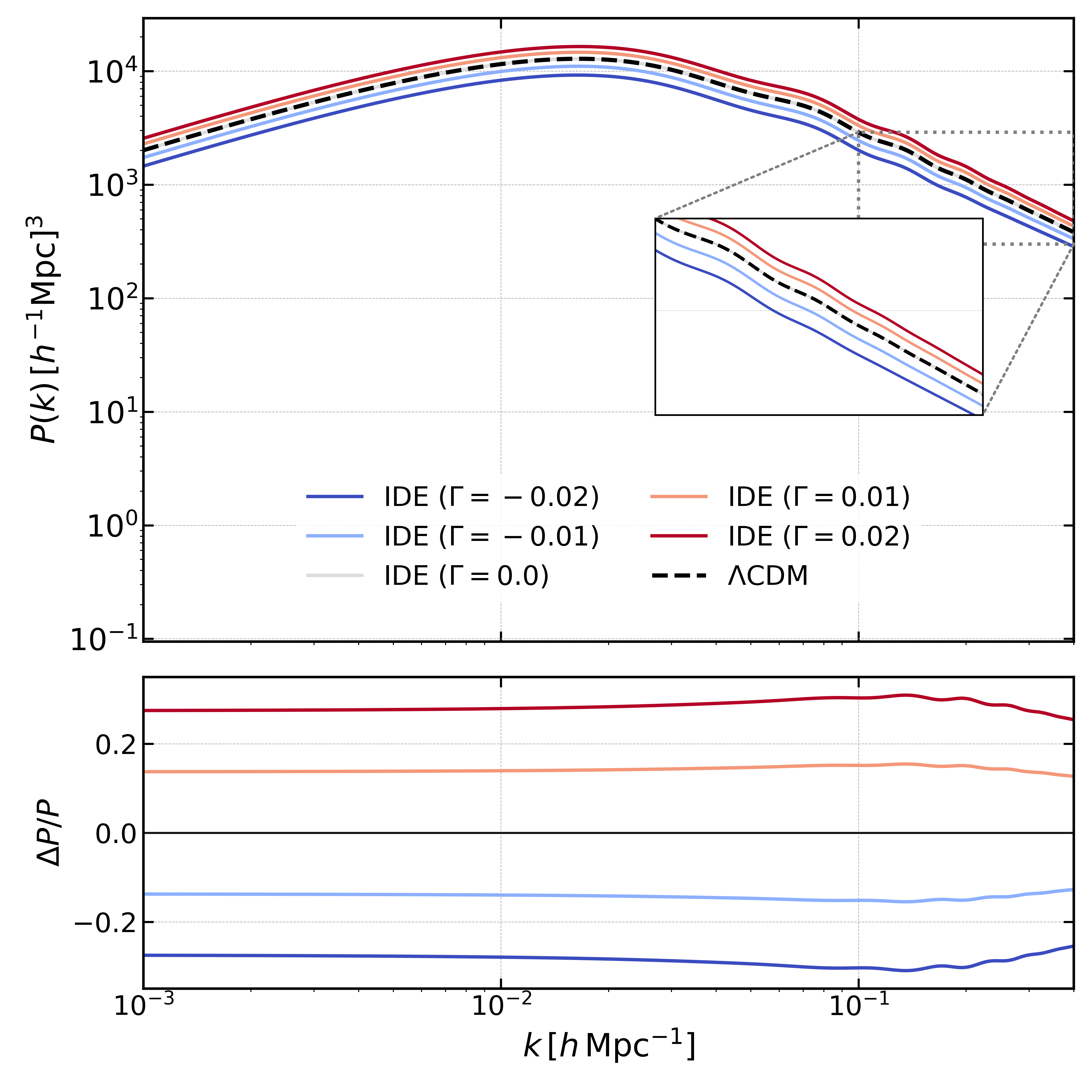}
    \includegraphics[scale=0.36]{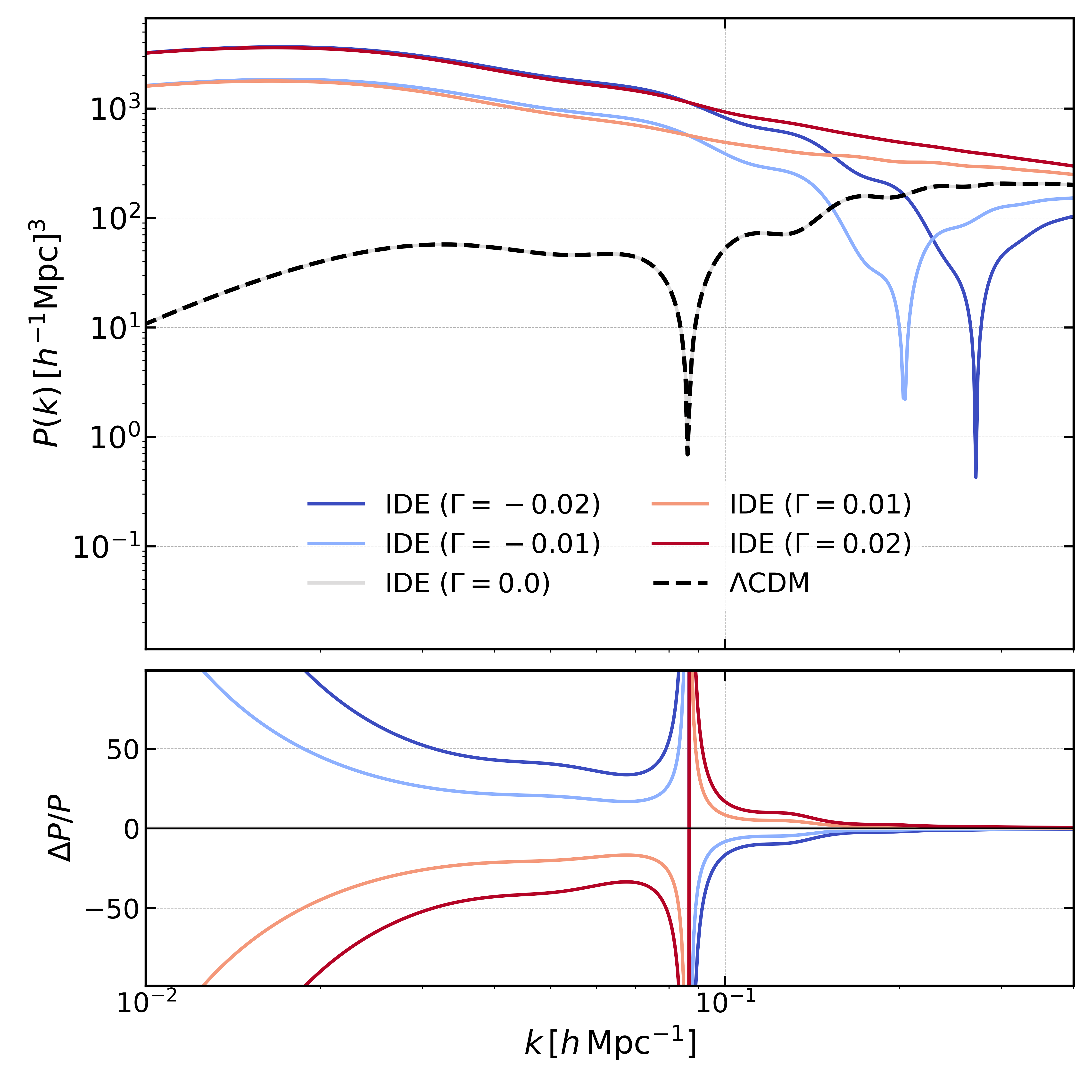}
    \caption{Left: Total matter power spectrum in real space within the IDE framework for different values of $\Gamma$, with the $\Lambda$CDM result shown as a black dashed curve. Right: total one-loop contributions for various $\Gamma$ values in the IDE model, with the $\Lambda$CDM prediction again indicated by the black dashed curve. In both panels we show the relative difference, where in the limit $\Gamma = 0$ the IDE model reduces to $\Lambda$CDM.}
    \label{fig:Pk_loop_IDE}
\end{figure}

For a first qualitative assessment, we plot the matter power spectrum including the perturbative corrections derived above. This is shown in Fig.~\ref{fig:Pk_loop_IDE}, which highlights the impact of the IDE-induced one-loop contributions on both the amplitude and scale dependence of the matter power spectrum. Varying the nonlinear coupling parameter introduced in Eq.~\eqref{eq:paramet_new} reveals a clear perturbative trend: positive values enhance the magnitude of the loop corrections, while negative values suppress them. As expected, when $\alpha_{\rm IDE}(\tau)=0$, all IDE-driven contributions vanish identically and the prediction reduces exactly to standard SPT. Since these modifications arise purely at the perturbative level, their influence manifests primarily as moderate shifts in the amplitude of the loop terms, without altering any of the physical scales set by the background expansion. In particular, perturbative corrections lack the capacity to modify the matter–radiation equality scale, $k_{\rm eq} = a_{\rm eq} H_{\rm eq}$, and therefore cannot shift the turnover position of the linear matter power spectrum. 

It is important to emphasize that, when making predictions to be compared with observations, all relevant quantities must ultimately be mapped into redshift space. At this stage, it is crucial to specify precisely which components of the theoretical modeling are modified in the presence of IDE and how these modifications propagate into the redshift-space description.

\begin{itemize}
    \item \textit{Growth rate in redshift-space kernels.} All redshift-space kernels (see ~\cite{Chudaykin:2020aoj} for instance) explicitly depend on the growth function \(f(a)\). Therefore, for accurate numerical evaluations, it is crucial to account for the fact that IDE models generally predict modifications to the growth function. In particular, one should assume
    \begin{equation}
        f(a) = \frac{d \ln D(a)}{d \ln a},
    \end{equation}
    where \(D(a)\) is given by Eq. (\ref{D_equation}). This ensures that the modified growth rate is consistently incorporated into all redshift-space calculations.

    \item \textit{Dependence on the linear power spectrum.} Both the integrals defining \(P_{\rm 1-loop, SPT}(z, k)\) and the leading counterterm contributions \(P_{\rm ctr}(z, k)\) explicitly depend on the linear power spectrum \(P_L(z,k)\). In IDE cosmologies, the linear power spectrum is highly sensitive to the model-specific predictions, as extensively discussed in the literature. Consequently, the numerical inputs for \(P_L(z,k)\) must be adjusted according to the interaction function \(Q\). In other words, \(P_L(z,k)\) is no longer simply the \(\Lambda\)CDM linear spectrum; it also depends on the underlying IDE coupling and its effect on linear growth.

    \item \textit{Impact on redshift-space kernels through \(F_2\) and \(F_3\).} In addition to the growth rate \(f(a)\), the redshift-space kernels \(K_2\) and \(K_3\) depend explicitly on the real-space perturbation-theory kernels \(F_2\) and \(F_3\), which receive IDE-induced corrections. However, in the regime where \(\Gamma \approx 10^{-3}\), the additional factors introduced by the IDE parametrization—such as \(\Gamma\,\rho_{\rm DE}\)—remain numerically small. As a consequence, the impact of these terms on the redshift-space kernels is effectively degenerate with the standard EFT bias operators. In practice, such contributions can be absorbed into redefined effective bias parameters (e.g., \(b_1\), \(b_2\)), without modifying the structure of the perturbative model. This ensures that IDE corrections remain well controlled while preserving a stable and tractable description in redshift space.
\end{itemize}

All derivations and approximations presented above are theoretically well motivated for investigating IDE scenarios within this perturbative framework. To the best of our knowledge, this is the first work in the literature to explore the impact of IDE on the perturbative treatment of structure formation at this level of detail. A systematic investigation of additional effects, together with a complete treatment in redshift space, lies beyond the scope of the present analysis and is deferred to future work. Since the interaction parameter \(\Gamma\) is expected to be close to zero, no significant deviations from the standard perturbative structure are anticipated, and the approximations adopted here remain robust for a first investigation of these effects. In the following section, we derive updated observational constraints using the theoretical setup and the perturbative IDE methodology developed above.

\section{Observational Constraints}
\label{sec:constraints_IDE}

To test our framework against observations, we employ the Einstein-Boltzmann solver \texttt{CLASS-PT}~\cite{Chudaykin:2020aoj}, a code specifically designed to incorporate perturbative corrections on mildly nonlinear scales. We interface this solver with the \texttt{MontePython} sampler~\cite{Audren:2012wb, Brinckmann:2018cvx}, which performs parameter inference using Markov chain Monte Carlo (MCMC) methods. Within this setup, we use the publicly available FS likelihood\footnote{Available at \url{https://github.com/oliverphilcox/full_shape_likelihoods/tree/main}.} based on BOSS DR12 data~\cite{Ivanov:2019pdj}. In this context, we implement the necessary perturbative modifications required to consistently incorporate the IDE dynamics into the FS modeling.

Convergence of the MCMC chains was ensured using the Gelman–Rubin criterion~\cite{Gelman_1992}, requiring $R - 1 \leq 10^{-2}$. The cosmological parameters sampled were the physical cold DM density $\omega_{\rm c} = \Omega_{\rm c}h^2$, the Hubble parameter $H_0$, the amplitude of the primordial scalar power spectrum $A_{\mathrm{s}}$, and the nonlinear coupling parameter $\Gamma$. Since BOSS data provide limited sensitivity to the spectral tilt, we chose not to vary it in our MCMC chains and instead fixed $n_{\mathrm{s}} = 0.9649$, as measured by \textit{Planck}~\cite{Planck:2018vyg}. Uniform priors were adopted over the following ranges:
\begin{align*}
    \omega_{\rm c} \in [0.0, 1.0], \quad H_0 \in [0.0, 100], \quad \ln(10^{10}A_{\mathrm{s}}) \in [0.0, 10.0], \quad \Gamma \in [-1.0, 1.0].
\end{align*}

Additional EFTofLSS parameters describing galaxy bias were included: the linear bias $b_1$, the quadratic bias $b_2$, and the tidal bias $b_{G_2}$, each independently sampled for the redshift bins considered. Postprocessing of the MCMC chains was performed using the \texttt{GetDist} package\footnote{Available at \url{https://github.com/cmbant/getdist}.}~\cite{Lewis:2019xzd}, from which we obtained the triangular plots as well as the tables summarizing the parameter constraints.

Our analysis uses the monopole and quadrupole ($\ell = 0, 2$) of the BOSS DR12 galaxy power spectrum~\cite{alam2015eleventh}. The data are split into two effective redshift bins, $z_{\rm eff}=0.38$ and $z_{\rm eff}=0.61$, each separated into NGC and SGC regions, yielding four independent subsamples. This configuration follows previous FS studies~\cite{Philcox:2021kcw, BOSS:2016psr, Ivanov:2019pdj} and combines CMASS and LOWZ galaxies. We employ the publicly available \texttt{CMASSLOWZTOT} catalog, including its window functions and random catalogs, to model survey geometry and selection effects. Covariance matrices are derived from 2048 \texttt{MultiDark-Patchy} mocks~\cite{Kitaura:2015uqa, Rodriguez-Torres:2015vqa}, calibrated to $N$-body simulations. We collectively refer to this dataset as \texttt{BOSS-DR12}, comprising the power-spectrum multipoles (P$_0$/P$_2$), the real-space proxy (Q$_0$), and the BAO rescaling parameters (AP). Additional details on the FS likelihood construction are provided in~\cite{Philcox:2021kcw}.

\begin{table}[htpb!]
\centering
\renewcommand{\arraystretch}{1.3}
\setlength{\tabcolsep}{12pt}
\begin{tabular}{lcc} \hline
\textbf{Parameter} & \textbf{IDE} & \textbf{$\Lambda$CDM} \\ \hline\hline
$\Omega_{\rm c} h^2$  & $0.1221\pm 0.0056$ & $0.1224\pm 0.0058$ \\
$\ln(10^{10} A_{s})$ & $2.81\pm 0.15$ & $2.85\pm 0.12$ \\
$H_0 \, [\mathrm{km/s/Mpc}]$ & $67.95\pm 0.82$ & $67.99\pm 0.85$ \\
$\Omega_{\rm m}$ & $0.3141\pm 0.0099$ & $0.314\pm 0.010$ \\
$S_8$ & $0.749\pm 0.054$ & $0.763^{+0.044}_{-0.049}$ \\ \hline
$\Gamma$ & $0.0039\pm 0.0082$ & -- \\ 
\hline\hline
\end{tabular}
\caption{Summary of the cosmological constraints for the IDE model. The quoted values correspond to the mean and 68\% CL, with the marginalized limits in parentheses.}
\label{table_IDE_main}
\end{table}

Table \ref{table_IDE_main} presents the cosmological constraints obtained for the IDE scenario using a FS analysis of the galaxy power spectrum. Importantly, these results rely exclusively on the shape information of the matter power spectrum inferred from galaxy clustering, without incorporating external datasets such as CMB, BAO, or supernova measurements. This approach allows us to investigate the constraining power of LSS data alone on IDE parameters, particularly the dark-sector coupling. In addition to the IDE model, in the last column of Table~\ref{table_IDE_main} we report, for comparison, the results obtained by running the analysis with the same priors but now applied to the $\Lambda$CDM model.

Within the IDE framework, we find that the physical cold DM density, $\Omega_{\rm c} h^2 = 0.1221 \pm 0.0056$, is tightly constrained by the power-spectrum shape information. This indicates that galaxy clustering retains substantial sensitivity to the matter content even in extended cosmological models. The amplitude of primordial scalar fluctuations, $\ln(10^{10}A_s) = 2.81 \pm 0.15$, is also reasonably well determined, though with the expected degradation relative to CMB-based analyses, reflecting the reliance on late-time large-scale-structure observables. The inferred Hubble constant, $H_0 = 67.95 \pm 0.82~\mathrm{km\,s^{-1}\,Mpc^{-1}}$, remains fully consistent with early universe $\Lambda$CDM determinations, suggesting that FS power-spectrum measurements alone contain sufficient information about the matter–radiation equality scale and the turnover to effectively constrain the expansion rate.

For the clustering parameter $S_8 = 0.749 \pm 0.054$, we observe a mildly lower value compared to standard CMB predictions and somewhat closer to several weak-lensing measurements, although current uncertainties preclude any definitive statement regarding the $S_8$ tension. Finally, the total matter density, $\Omega_{\rm m} = 0.3141 \pm 0.0099$, is strongly constrained by the power-spectrum shape alone, highlighting the robustness of FS analyses in determining both background and clustering properties in cosmological models beyond $\Lambda$CDM.

When comparing the IDE and $\Lambda$CDM columns, we find that incorporating only the one-loop perturbative corrections specific to the IDE framework does not lead to any statistically significant shifts in the inferred cosmological parameters relative to the standard model. This shows that, at the precision of BOSS-DR12 data, the IDE one-loop contributions do not materially affect the resulting parameter constraints.

\begin{figure}[htpb!]
    \centering
    \includegraphics[width=0.57\columnwidth]{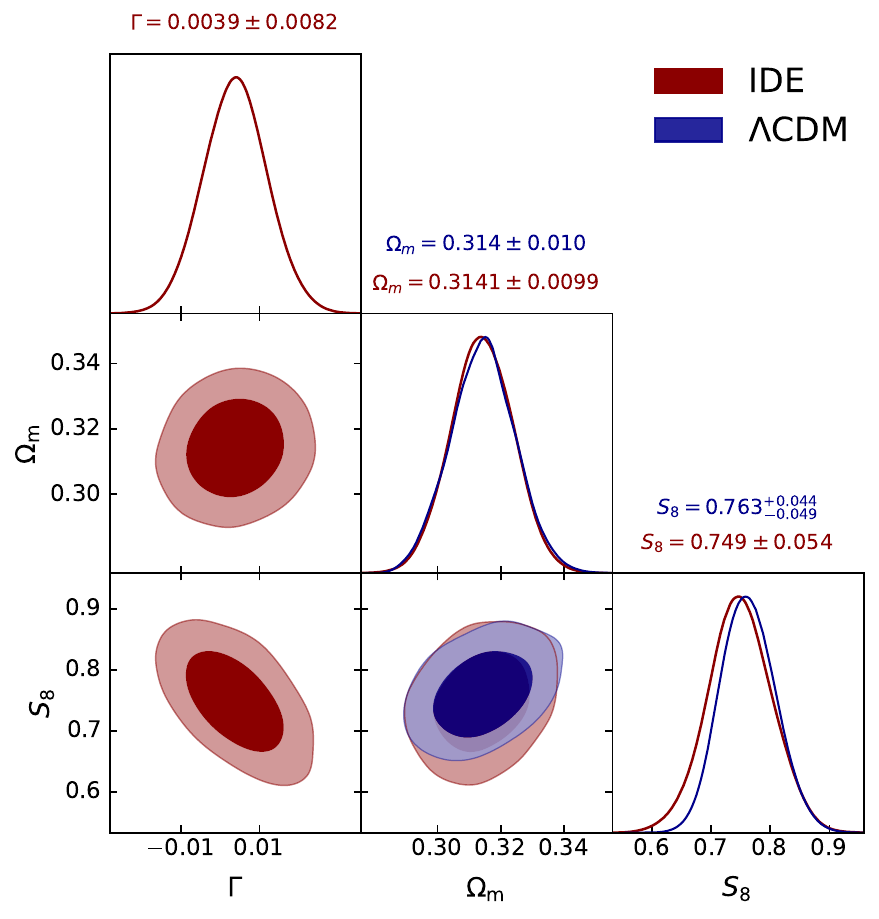} 
    \caption{Posterior distributions and 68\% and 95\% credible contours for the parameters $\Gamma$, $S_{8}$, and $\Omega_{\rm m}$. Blue corresponds to $\Lambda$CDM and red to IDE, both analyses using BOSS-DR12 data.}
    \label{PS_IDE_main}
\end{figure}

A central result for interpreting these findings is the constraint on the dark-sector interaction parameter, 
$\Gamma = 0.0039 \pm 0.0082$, which remains fully consistent with zero at the $1\sigma$ level, in agreement with $\Lambda$CDM. Obtaining such a bound using \emph{only} the shape of the galaxy power spectrum is particularly noteworthy. In interacting DE models, the coupling modifies the growth rate of cosmic structures, the scale dependence of growth on linear and quasilinear scales, the effective gravitational strength at late times, and the relative evolution of DM and DE densities. These effects imprint characteristic signatures on the shape of the matter power spectrum. The fact that we recover this level of constraint while incorporating only one-loop perturbative corrections highlights the value of FS analyses, beyond probes relying exclusively on BAO or distance information, for capturing the subtle, scale-dependent modifications induced by a nonzero interaction. Figure~\ref{PS_IDE_main} displays the marginalized $1\sigma$ and $2\sigma$ confidence regions for selected parameters of interest. Notably, introducing the coupling parameter $\Gamma$ generates a non-negligible negative correlation with $S_8$, consistent with expectations for IDE scenarios in which the interaction can appreciably alter the growth rate of cosmic structures.

The ability of galaxy power-spectrum data alone to constrain $\Gamma$ at the level of $\sim 10^{-2}$ highlights the strong sensitivity of LSS to deviations from standard gravity and dark-sector physics. While no evidence for a nonzero interaction is found, the allowed range for $\Gamma$ still permits a mild energy exchange between DM and DE. Future surveys with improved volume, redshift precision, and nonlinear modeling will significantly enhance this sensitivity, potentially reducing the uncertainty on $\Gamma$ by a factor of several.

\section{Final remakes}
\label{sec:end_IDE}

It is well known that mildly nonlinear regimes of structure formation contain a large amount of cosmological information. These regimes not only help strengthen parameter constraints but also serve as powerful complementary probes for testing departures from the standard cosmological model. Although their theoretical and computational treatment has historically been challenging, rapid progress in perturbative techniques, simulations, and effective descriptions has significantly advanced the field in recent years. In this context, FS analyses of galaxy clustering have become established as one of the most robust observational methods for jointly probing cosmological parameters, DE, and DM.

Motivated by this broader effort to explore observational signatures of models beyond the $\Lambda$CDM paradigm, IDE scenarios have attracted considerable attention in recent years. In this work, we develop the real-space perturbative calculations for a specific IDE parametrization and evaluate their impact within the EFTofLSS framework. As a first proof of concept, we show that corrections appearing solely at the level of the galaxy power spectrum, $P_g(k)$, already allow us to constrain the coupling parameter at the ${\cal O}(10^{-2})$ level using galaxy-only FS data. This demonstrates the strong constraining power of mildly nonlinear scales for this class of models. We find no statistically significant deviations from the $\Lambda$CDM cosmology. Nonetheless, the IDE-induced terms generate new correlations absent in the standard scenario, highlighting potential avenues for enhanced sensitivity in future analyses.

The present study should be viewed as an initial methodological step toward systematically incorporating physics beyond $\Lambda$CDM, particularly IDE models, into analyses of mildly nonlinear clustering. With the imminent release of public galaxy, quasar, and Lyman-$\alpha$ forest clustering measurements from the DESI~\cite{DESI:2024jxi, DESI:2024hhd} and from Euclid~\cite{Euclid:2025diy, Euclid:2023tog}, it will become possible to test these scenarios with unprecedented precision and to further tighten the limits on potential deviations from $\Lambda$CDM in the nonlinear regime. Furthermore, combining multiple datasets and modeling approximations with clustering measurements is essential for achieving a comprehensive and nontrivial characterization of these effects in interacting de scenarios.

\begin{acknowledgments}
We thank the referee for the careful reading of the manuscript and for the constructive comments and suggestions, which helped to improve the clarity and presentation of this work. E.S. received support from the CAPES scholarship. R.C.N. thanks the financial support from the Conselho Nacional de Desenvolvimento Científico e Tecnologico (CNPq, National Council for Scientific and Technological Development) under the project No. 304306/2022-3, and the Fundação de Amparo à Pesquisa do Estado do RS (FAPERGS, Research Support Foundation of the State of RS) for partial financial support under the project No. 23/2551-0000848-3.
\end{acknowledgments}

\appendix
\section{A NOTE ON THE CASE \( Q = \xi \mathcal{H} \rho_{\rm DE} \)}
\label{app:A}
In this Appendix, we present the equations that modify the $\delta\delta$ one-loop power spectrum in one of the most commonly used parametrizations for $Q$ in the literature, namely $Q = \xi \mathcal{H} \rho_{\rm DE}$. Following the same procedure adopted for the main results in this paper and defining $P_{22}$ and $P_{13}$ as the standard contributions to the one-loop term, we can express the corresponding equations in this IDE scenario by neglecting higher-order terms in $\xi$
\begin{equation}
    P_{\rm 1-loop}(k) = P_{22}(k) + P_{13}(k) + \Delta P_{22}(k) + \Delta P_{13}(k) 
\end{equation}
where
\begin{equation}
    \Delta P_{22}(k) =  - \frac{10}{7} \frac{\xi\rho_{DE}}{\rho_m} P_{22} ,
\end{equation}
and
\begin{equation}
\Delta P_{13}(k)
= 6\,P_L(k)\int d^3q \;
\Delta F_3(\mathbf{k},\mathbf{q},-\mathbf{q}) \; P_L(q),
\end{equation}
with
\begin{equation}
\begin{split}
\Delta F_3(\mathbf{q_1,q_2,q_3}) 
= \frac{\lambda}{18 + 7\lambda} \Big[
    -5\,\alpha(\mathbf{q_1,q_{23}}) F_2(\mathbf{q_2,q_3})
    - \frac{10}{7}\,\beta(\mathbf{q_1,q_{23}}) G_2(\mathbf{q_2,q_3})
    + \frac{4}{7}\,\beta(\mathbf{q_1,q_{23}})\beta(\mathbf{q_2,q_3}) \\
\qquad
    -5\,G_2(\mathbf{q_1,q_2}) \alpha(\mathbf{q_{12},q_3})
    - \frac{10}{7}\,\beta(\mathbf{q_{12},q_3}) G_2(\mathbf{q_1,q_2}) \\
\qquad
    + 2\,\beta(\mathbf{q_1,q_2})\alpha(\mathbf{q_{12},q_3})
    + \frac{4}{7}\,\beta(\mathbf{q_{12},q_3})\beta(\mathbf{q_1,q_2})
\Big].
\end{split}
\end{equation}

Here we defined 
\[
\lambda(a) = \frac{\xi\, \rho_{\mathrm{de}}(a)}{\rho_m(a)},
\]
and when $\xi \to 0$ we recover the standard (noninteracting) theory. This result highlights another important feature: within the IDE framework, the power spectrum is highly sensitive to the specific parametrization of $Q$. In the nonlinear regime, while the choice $Q = \xi \mathcal{H}\rho_{\rm DM}$ does not affect the growth of structures, the alternative parametrization $Q = \xi \mathcal{H}\rho_{\rm DE}$ introduces additional terms, $\Delta P_{22}(k)$ and $\Delta P_{13}(k)$, which modify the full one-loop contribution. The model $Q = \xi H \rho_{\rm DE}$ was recently studied in \cite{PRL_MIG}, for example. One should note that this parametrization can affect cosmology on multiple fronts: background evolution, linear perturbations impacting $P_L(k)$ and $D(z)$, as well as nonlinear scales, among other effects. A comprehensive investigation of these impacts, and their implications for a broad set of observables such as CMB, BAO, SNe Ia, galaxy clustering, and cosmic shear, requires more careful and exhaustive analyses, and will be presented in future work.

\bibliographystyle{apsrev4-1}
\bibliography{main}

\end{document}